\documentclass[a4paper,11pt,headings=big,DIV=12]{scrartcl}
\pdfoutput=1
\usepackage{amsmath,amssymb,mathtools}
\usepackage[colorlinks,linkcolor=mygray,urlcolor=blue,citecolor=blue]{hyperref}
\usepackage{tcolorbox} 
\definecolor{mygray}{gray}{0.2}
 
\definecolor{mypink1}{rgb}{0.9, 0.2, 0.6}
\usepackage{graphicx}
\addtokomafont{disposition}{\rmfamily\boldmath}
\usepackage[utf8]{inputenc}
\usepackage{enumerate} 
\usepackage{slashed}
\usepackage{cite} 
\usepackage{comment}
\usepackage{multirow}
\usepackage{acronym}
 \usepackage{bbold}
\allowdisplaybreaks
\acrodef{PDG}[PDG]{Particle Data Group}
\acrodef{OPE}[OPE]{Operator Product Expansion}
\acrodef{FCNC}[FCNC]{flavour-changing neutral current}
\acrodef{RHC}[RHC]{right-handed currents}
\acrodef{SM}[SM]{Standard Model}
\acrodef{NP}[NP]{New Physics}
\acrodef{MFV}[MFV]{Minimal Flavour Violation}
\acrodef{SD}[SD]{short-distance}
\acrodef{LD}[LD]{long-distance}
\acrodef{DA}[DA]{distribution amplitude}

\newcommand{\CondQQ}[1]{{ \left\langle \bar{#1} #1 \right \rangle}}
\newcommand{\CondQGQ}[1]{{ \left \langle \bar{ #1 } \sigma \! \cdot \! G #1 \right \rangle }}
\newcommand{\Condqqff}[1]{ \vev{ V^a_{#1}  V^a_{f} } }

\newcommand{\vev}[1]{\langle #1 \rangle}

\newcommand{\matel}[3]{\langle #1|#2|#3\rangle}
\newcommand{\al}{\alpha}
\newcommand{\be}{\beta}
\newcommand{\ga}{\gamma}
\newcommand{\de}{\delta}
\newcommand{\la}{\lambda}
\newcommand{\eps}{\epsilon}
\newcommand{\veps}{\varepsilon}

\newcommand{\GeV}{\,\mbox{GeV}}
\newcommand{\MeV}{\,\mbox{MeV}}

\newcommand{\Tr}{{\textrm{Tr}}}

\newcommand{\Dlra} {\overset{\leftrightarrow}{D}}

\newcommand{\amp}{ \mathcal A}
\newcommand{\eff}{\textrm{eff}}
\newcommand{\SM}{\textrm{SM}}
\newcommand{\NP}{\textrm{NP}}

\newcommand{\SU}{\text{SU}}

\newcommand{\U}{\text{U}}

\newcommand{\VmA}{}
\newcommand{\VpA}{'}
\newcommand{\ampt}{A}

\newcommand{\Vf}{V \ga}
\newcommand{\DRHC}{\Delta_R e^{i \phi_{\Delta_R}} }

\newcommand{\fone}{f_1}
\newcommand{\foneone}{f_1(\text{\small{1285}})}
\newcommand{\fonetwo}{f_1(\text{\small{1420}})}
\newcommand{\hone}{h_1}
\newcommand{\honeone}{h_1(\text{\small{1170}})}
\newcommand{\honetwo}{h_1(\text{\small{1380}})}
 
\newcommand{\Koneone}{K_1(\text{\small{1270}})}
\newcommand{\Konetwo}{K_1(\text{\small{1400}})}

\newcommand{\DD}{D} 
\newcommand{\UU}{U}
\newcommand{\HH}{H}

\newcommand{\RR}{\mathbb{R}}
\newcommand{\GG}{G}
\newcommand{\epsQCD}[1]{{\eps}^{#1}}
\newcommand{\epsQCDp}[1]{{\eps}^{'#1}}
\newcommand{\laCKMn}[1]{\tilde{\la}_{#1}}
\newcommand{\Rea}{\textrm{Re}}
\newcommand{\Ima}{\textrm{Im}}
\newcommand{\sincos}{\!\! \begin{array}{c} \sin \\[-0.3cm] \cos \end{array}\!\!}

\newcommand*{\mathcolor}{}
\def\mathcolor#1#{\mathcoloraux{#1}}
\newcommand*{\mathcoloraux}[3]{%
  \protect\leavevmode
  \begingroup
    \color#1{#2}#3%
  \endgroup
}

\begin{document}

\begin{flushright}
\begin{tabular}{l}
CP3-Origins-2018-13 DNRF90
\end{tabular}
\end{flushright}
\vskip1.5cm

\begin{center}
{\Large\bfseries \boldmath Parity Doubling 
as a Tool for \\[0.1cm]
Right-handed Current Searches}
\\[0.8 cm]
{\Large%
 James Gratrex and Roman Zwicky
\\[0.5 cm]
\small
 Higgs Centre for Theoretical Physics, School of Physics and Astronomy,\\
University of Edinburgh, Edinburgh EH9 3JZ, Scotland 
} \\[0.5 cm]
\small
E-Mail:
\texttt{\href{mailto:roman.zwicky@ed.ac.uk}{roman.zwicky@ed.ac.uk} ,
\href{mailto:j.gratrex@ed.ac.uk}{j.gratrex@ed.ac.uk}}.
\end{center}

\bigskip
\pagestyle{empty}

\begin{abstract}\noindent
The V-A structure of the weak interactions leads to definite amplitude hierarchies 
in exclusive heavy-to-light decays mediated by $b \to (d,s)\gamma$ and $b \to (d,s) \ell \bar{\ell}$. 
However,  the extraction of right-handed currents beyond the Standard Model
 is contaminated by V-A long-distance contributions leaking into right-handed amplitudes. 
We propose that these quantum-number changing long-distance contributions can be 
controlled by considering the almost parity-degenerate vector meson final states
by exploiting the \emph{opposite} relative sign of left- versus right-handed amplitudes.
For example, measuring the time-dependent rates of a pair of vector $V(J^P=1^-)$ and axial $A(1^+)$ mesons
in $B \to (V,A) \gamma$, up to an order of magnitude is gained on the theory uncertainty prediction, 
controlled by long-distance ratios  to the right-handed amplitude. 
This  renders these decays 
clean probes to null tests, from the theory side.
 \end{abstract}

\newpage

\setcounter{tocdepth}{3}
\setcounter{page}{1}
\pagestyle{plain}

\tableofcontents

\section{Introduction}

It is well-known that the V-A  structure of the weak interaction leaves characteristic traces 
in the polarisation of weak decays e.g.
 \cite{KG79,AGS97,BR05}.  This is particularly attractive in heavy-to-light \acp{FCNC},
such as $b \to \DD \ga$ or $b \to \DD \ell \bar{\ell}$ with $\DD = d,s$ and $\ell = e,\mu,\tau$. 
These transitions therefore make 
excellent probes to search for new physics with V+A structure, 
 referred to as \ac{RHC} below. 
 Schematically,  the effective Hamiltonian for such decays reads
 \begin{equation}
 \label{eq:Heff1}
H_{\eff}^{b \to d \ga} \sim C \,  \bar d_L \Gamma b \, O_r   +  C' \,  \bar d_R \Gamma b \, O_r'   \;,  \quad 
C^{(')} = C^{(')}_\SM + C^{(')}_\NP \;,
\end{equation} 
 where $2d_{L(R)} = (1 \mp \ga_5) d$, and  $O^{(')}_r$ stands for the (r)emaining part of the operator, 
 to be made more precise in section \ref{sec:main}. The $b \to s \ga$ case is recovered by $d \to s$,  with  
 according changes in the CKM factors. 
 In the \ac{SM}, $(C'/C)_\SM \sim m_\DD/m_b$, whereas in generic \ac{NP} models 
 $(C'/C)_\NP \gg (C'/C)_\SM$ \cite{Agashe:2003rj,Lunghi:2006hc,Kou:2013gna,Konig:2014iqa}, with 
 current constraints on the electroweak penguin operator around $(C_7'/C_7)_\NP \sim 1/5$ 
 \cite{Descotes-Genon:2015uva,Paul:2016urs,Hurth:2017hxg}. 
 In the terminology of the \ac{MFV} effective field theory framework, 
 the $O$- and $O'$-type operators, with the above-mentioned hierarchies,  are referred to as \ac{MFV}  and non-\ac{MFV} type \cite{MFV}. 
 
Non-perturbative matrix elements, connecting  $H_{\eff}$  to amplitudes, can dilute the cleanliness of 
the signal.  For $\bar{B} \to V \gamma$, where $V$ is a vector meson (e.g. $\rho, K^*, \dots$), the form factors,
 referred to as \ac{SD} contributions hereafter,  
obey exact algebraic relations, leading to accidental control 
in the \ac{SD} part.  
However, sizeable tree-level four-quark operators with charm and up quarks,  
$H_{\eff} \sim \bar d_L \ga_\mu \UU\bar \UU_L \ga^\mu b$ ($\UU = u,c$), induce genuine \ac{LD} effects, which 
are more difficult to control.  It was argued, based on studying the inclusive $\bar{B} \to X_{s} \ga$ decay,
that such contaminations could be rather significant  \cite{GGLZ2004}, whereas 
actual computations show smaller effects in exclusive channels \cite{prep_charm,MXZ2008,BJZ2006,BZ06CP,KSW1995}.

In this article, we show that these \ac{LD} effects can be controlled by a symmetry that in turn 
also explains the smallness found in the concrete computation \cite{prep_charm} quoted just above.
The symmetry in question is the \emph{chiral restoration limit}.  
The crucial point is that decays of 
opposite parity, such as  $\bar{B} \to \rho(1^{--})  \ga$ versus $\bar{B} \to a_1(1^{++})  \ga$, 
are opposite in sign in the right-handed amplitude between the
exact \ac{SD} and  \ac{LD} contributions (originating from the sizeable V-A part).\footnote{We choose to use the $\rho$ and $a_1$ as the prime examples for general 
 discussions on historical grounds, in connection with the Weinberg sum rules \cite{WSR,Weinberg2}. Other parity doubling pairs, with considerably smaller widths, are tabulated in section \ref{sec:doubleQCD}. 
 For an exhaustive review on the physics and history of parity doubling, we refer the reader to reference
 \cite{Afonin:2007mj}. Some more discussion can found in appendix \ref{app:PD}. 
 } 
While decays of axial mesons have received some attention as complementary probes for \ac{RHC} (e.g. \cite{Gronau:2001ng,BaBar2015CP,Kou:2016iau}), we advocate that the combination of the two decay channels allows for a cleaner extraction of the relevant observables  controlled by ratios of vector to axial \ac{LD} amplitudes. 
In light-cone approaches, this necessitates axial vector meson \acp{DA} \cite{Yang07}, whose 
symmetry relations with vector meson \acp{DA} can be studied rather systematically \cite{prep_da}.
For a simplified discussion of the main ideas of this paper we refer the reader to \cite{Gratrex:2018bhz}.

The paper is organised as follows. In section \ref{sec:main} it is shown, using
the path integral, 
that the fraction of \ac{LD}- over \ac{SD}-\ac{RHC} flips sign for parity doublers.  
In section \ref{sec:beyond}, the parity doublers are listed (section \ref{sec:doubleQCD}), followed 
by a discussion of the sources of correction to the symmetry limit in section \ref{sec:ratios}. 
Applications to the  time-dependent rates of $\bar{B} \to (V,A) \ga$, a detailed breakdown of
$\bar{B}_s \to \phi(f_1) \ga$, and remarks on  $B \to (V,A) \ell \bar{\ell}$ are presented 
in sections  \ref{sec:TDCP},  \ref{sec:exemplified} and  \ref{sec:other}  respectively.
 The paper ends with 
conclusions in section \ref{sec:conc}, including comments on the experimental feasibility of the measurement. 
A discussion on the chiral order parameter, illustrating some technical aspects of the paper, is given in appendix \ref{app:pi}. 

\section{The use of parity doubling for right-handed current searches}
\label{sec:main}

After briefly discussing the structure of  $B \to V \ga$ amplitudes  in section  \ref{sec:ChiralityDef},
we demonstrate in section  \ref{sec:Restoration} how the left- and right-handed amplitudes 
of opposite parity states come with a relative minus sign.

\subsection{Chirality hierarchy of amplitudes in the Standard Model}
\label{sec:ChiralityDef}

The $B \to V \ga$ amplitude can be expressed 
in terms of the two photon polarisations as\footnote{The extension to the notation of $B \to V \ell \bar{\ell}$ is as follows: 
$\amp_{L(R)} \sim H_{\mp}$, with photon polarisation vectors $\veps(\pm) = (0,\pm 1 ,i,0)/\sqrt{2}$, 
and the amplitudes are frequently written in terms of $\sqrt{2} \amp_{\perp( \parallel)}  = H_+ \mp H_- $ \cite{HZ2013,GHZ2016}. Away from $q^2=0$, one also needs the amplitude $\amp_0$, corresponding to the longitudinal polarisation of 
the vector meson or the off-shell photon.} 
\begin{equation}
\label{eq:amp-pol}
\amp \equiv \matel{\ga(q,\veps) V(p,\eta)  }{ H_\eff}{\bar{B}(p_B)}
 = \bar{\amp}^{\bar{B} \to \Vf}_L  S_L + \bar{\amp}^{\bar{B} \to \Vf}_R S_R \;,
\end{equation}
where $S_{L(R)} \equiv [\epsilon(\veps^*, \eta^*, p ,q) \pm i \{ (\veps^* \eta^*)(pq) - (\veps^* p) ( \eta^*q)   \}]$, 
Levi-Civita sign convention $\eps_{0123} = 1$, contractions of vectors are understood,
and $\veps$ and $\eta$ are the polarisation vectors of the photon and the meson respectively.
The bar refers to $\bar{B}$ transitions ($b \to D \ga$), where $D=(d,s)$, as opposed to the $B$-transition ($\bar{b} \to \bar{D} \ga$).
Each chirality amplitude can then be decomposed into contributions from $O$ 
and $O'$ operators \eqref{eq:Heff1}:
\begin{equation}
\label{eq:dec}
\bar{\amp}^{\bar{B} \to \Vf}_\chi = \bar{\ampt}_\chi^{\VmA} + \bar{\ampt}_\chi^{\VpA}    \;, \quad \chi = L,R \;,
\end{equation}
dropping the superscript for brevity. The V-A interactions imply 
$| \bar{\ampt}_\chi^{\VmA} | \gg |\bar{\ampt}_\chi^{\VpA}|$, which is, for example, 
encoded in $C_7'/C_7 = m_D/m_b$ in the \ac{SM}, where the $C_7$ and $C_7'$ are Wilson coefficients 
of the effective Hamiltonian 
($\UU= u,c$ and $\lambda_{\UU }^{(\DD)} = V_{ \UU b}^{\vphantom{*}}V_{\UU \DD}^*$)
\begin{equation}
\label{eq:Heff}
H_{\eff}^{b \to (\DD=d,s) \ga}=  \frac{4 G_F}{\sqrt{2}}\Big(  \lambda_{\UU}^{(\DD)} 
\Big[ C_1 O_1^{\UU} + C_2 O_2^{\UU} \Big]  
-    \lambda_{t}^{(\DD)} \sum_{i=3}^8 C_i O_i    \Big) + \{ C,\DD_L \to C',\DD_R \} \;,
\end{equation}
with more detailed definitions in appendix \ref{app:effectiveH}.

Normalising to the dominant \ac{SD} contribution, the amplitudes \eqref{eq:dec}  read
\begin{alignat}{2}
 &  \bar{\amp}^{\bar{B} \to \Vf}_L &\;=\;&   4 \sqrt{2} G_F   \la_t C_7 T_1(0) \left(   1  +  
 \laCKMn{i}(   \epsQCD{i}_{V,L} + \epsQCDp{i}_{V,L}) \right)  \;, \nonumber \\[0.1cm]
 &  \bar{\amp}^{\bar{B} \to \Vf}_R &\;=\;&   4 \sqrt{2} G_F   \la_t C_7 T_1(0)  \left(  \hat{C}_7' +  
 \laCKMn{i}(   \epsQCD{i}_{V,R} + \epsQCDp{i}_{V,R}) \right)  \;, \quad 
\end{alignat}
with  summation over  $i = u,c$  implied, 
$- 2 T_1(0) S_{L(R)} = \matel{V}{ \bar s_{L(R)} \sigma \!\cdot\! F b}{\bar{B}} $,
and  $\epsQCD{i}_{V,\chi}$ includes the ratio of Wilson coefficients and the QCD matrix element but not 
the CKM contribution $\laCKMn{i}  \equiv \la_{i}/\la_{t}$ (with $(\DD)$ superscript suppressed).   
The \ac{NP} part $\DRHC$ of the \ac{RHC} is encoded in
\begin{equation}
\label{eq:C7pC7}
\hat{C_7'} \equiv \frac{C_7'}{C_7}   = \hat{m}_{d,s} + \DRHC \;, \quad \hat{m}_{\DD} \equiv m_\DD/m_b \;,
\end{equation}
where, by convention,  $\Delta_R   \geq 0$, and $\phi_{\Delta_R}$ is the weak (CP-odd) phase relative to the \ac{SM} phase originating from $\la^{(\DD)}_t$.  For further discussion, it is convenient to break down the relative parts into the following table:
\begin{equation}
\label{eq:tab-break1}
 \begin{tabular}{ l  |  l  l  |  l  l  }
$\bar{ \amp}^{\bar{B} \to \Vf}_\chi$  & $\bar{ \ampt}^{\VmA} _{SD,\chi} $         
  & $ \bar{ \ampt}^{\VmA} _{LD,\chi}$  
    &   $\bar{\ampt}^{\VpA} _{SD,\chi}$ & 
  $\bar{ \ampt}^{\VpA} _{LD,\chi} $ \\[0.2cm]
  \hline 
$\chi = L$ & $1$  &  $\laCKMn{i}\epsQCD{i}_{V,L}$ 
& $0$  & $\laCKMn{i}\epsQCDp{i}_{V,L}$ \\
$\chi = R$  & $0$ &  $\laCKMn{i}\epsQCD{i}_{V,R}$
& $\hat{m}_{d,s} + \DRHC $ 
&  $\laCKMn{i}\epsQCDp{i}_{V,R}$  
\end{tabular}  \;.
\end{equation}
The  two zero entries in \eqref{eq:tab-break1} are due to the algebraic relation 
$\sigma^{\al\be} \ga_5 = -  \frac{i}{2} \eps^{\al\be\ga\de}\sigma_{\ga\de}$,
 which descends to the form-factor relation  $T_1(0) = T_2(0)$.
 
The relative importance of the \ac{LD} contributions in $b \to D \ga$ depends on the CKM hierarchy \eqref{eq:Wolf}.
More specifically, the $\bar{ \ampt}^{\VpA} _{LD,\chi} $ are not of major importance, as 
only the $C'_8 O'_8$-operator contributes, and it was shown in \cite{DLZ2012} that, at leading twist, 
$\bar{ \ampt}^{\VpA} _{LD,L} =0  $, while
 $\bar{ \ampt}^{\VpA} _{LD,R} $ is at the percent level in the normalisation above.
The $\epsQCD{i}_{V,L,R}$ are the, potentially sizeable, \ac{LD} contributions. 
Throughout this presentation we assume $\epsQCD{i}_{V,L,R} \ll 1$, which is a circumstance 
that can be checked experimentally for the $\epsQCD{i}_{V,R}$ contribution (cf. section \ref{sec:exemplified}).

Crucially,  the breakdown \eqref{eq:tab-break1} reveals that, in a vector-meson final state of definite parity, $\epsQCD{i}_{V,R}$ cannot be distinguished from the \ac{RHC} $\DRHC$.
It is the aim of this work to show, however, that 
$\DRHC$ \emph{can} be unambiguously identified when two parity-doubler vector meson final states, to be listed 
in section \ref{sec:doubleQCD}, are taken into account. In order to gain some insight, we first discuss 
the procedure in the chiral symmetry restoration limit, before returning to QCD in section \ref{sec:beyond}.

\subsection{The chiral symmetry restoration limit}
\label{sec:Restoration}

We consider the effect of the chiral symmetry restoration limit on the breakdown 
\eqref{eq:tab-break1}, using $B \to \rho$ versus $B \to a_1$ as a template. 
In  this 
limit, suppressing the Baryon number $\U(1)_V$, 
the global flavour symmetry of $N_F$ fermions
is restored:
\begin{eqnarray}
\label{eq:restore}
& & \{  m_q , \vev{\bar q q} ,  \dots\}   \to 0  \;, \quad  \Rightarrow  
\quad  \SU(N_F)_V \to \SU(N_F)_V \times \SU(N_F)_A \times \U(1)_A \;,
\end{eqnarray}
where $ \SU(N_F)_V \times \SU(N_F)_A \simeq  \SU(N_F)_L \times \SU(N_F)_R$, and
the dots stand for other $\SU(N_F)_A\times \U(1)_A$-violating condensates such as
$\CondQGQ{q}$ (see appendix \ref{app:pi} for further 
discussion). Let us mention in passing that such a situation can be simulated on the lattice at temperatures 
above the chiral phase transition, cf. footnote \ref{foot:lattice} in appendix \ref{app:PD}.

\subsubsection{Path integral representation of matrix elements}
\label{sec:path}

\begin{figure}[h!]
\centering
\includegraphics[scale=0.7,clip=true,trim=20 520 0 70]{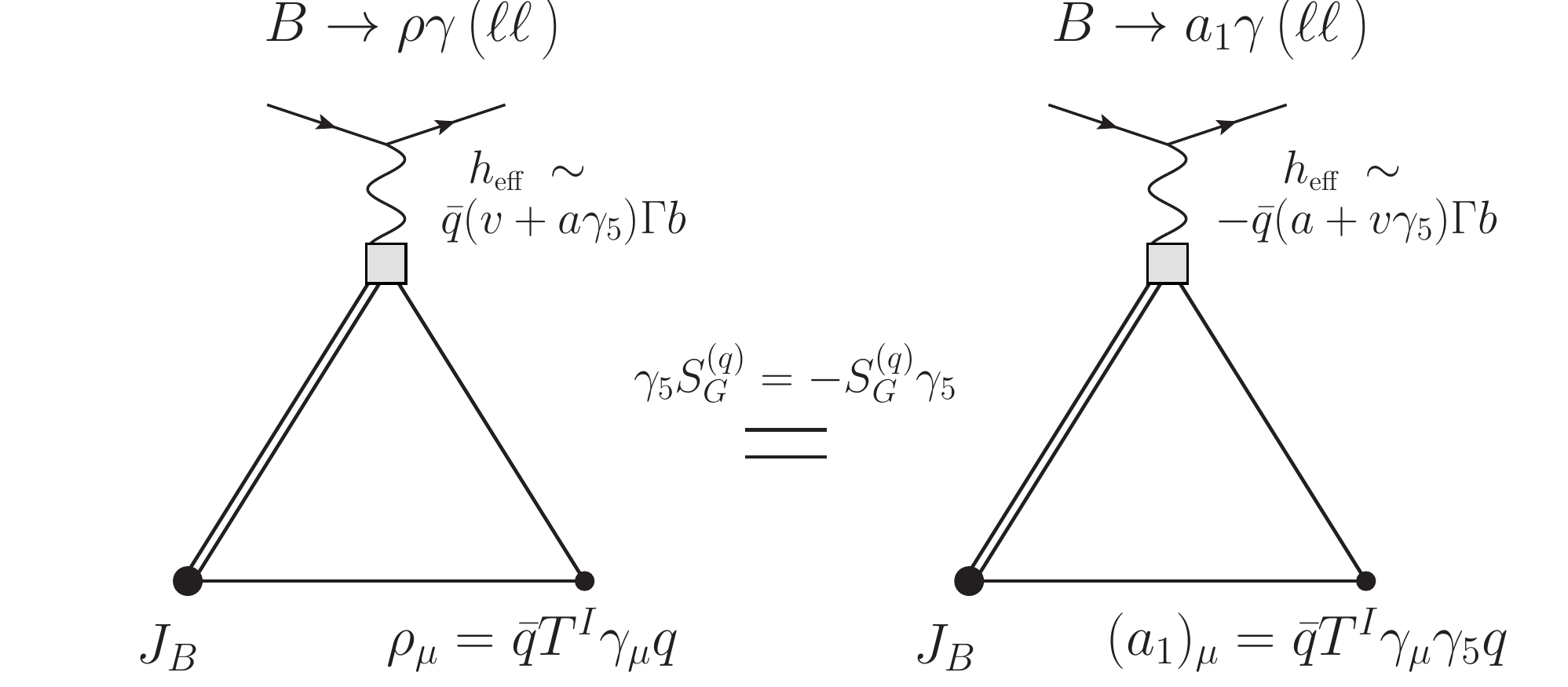}
\caption{\small Diagram representing the procedure outlined in the text, using the relation \eqref{eq:ga5}, which necessitates both the limits $m_q \to 0$ and $\CondQQ{q} \to 0$ in \eqref{eq:restore}. 
The argument only requires that
the weak vertex $h_{\textrm{eff}}$ be a local operator, 
and thus applies to both \ac{SD}  (form-factor) and \ac{LD} (charm-loop) contributions.  
Note that the argument also applies to the annihilation topology where the photon is emitted 
from the $b$-quark or light-quark lines in the triangle graph.
The schematic correlation functions on the left and right are exactly equal, from where the information on the matrix elements 
can be assessed via the LSZ formalism or dispersion relations, when taking into account finite-width effects. Corrections to this exact equality, beyond the chiral symmetry limit, are discussed in section \ref{sec:beyond}.}
\label{fig:PD_Diagram}
\end{figure}

In establishing our main result,  the path-integral formalism, with quarks propagating in a background gluon field, will prove powerful.\footnote{The realisation in concrete computations, e.g. light-cone sum rules, is rather 
subtle, which is to do with the nature of the restoration limit \cite{prep_charm}.}
 The $\rho$- and $a_1$-meson can be represented by the  
interpolating currents, with isospin $I$, 
\begin{equation}
\label{eq:inter}
\rho_\mu^I = \bar q  \ga_\mu T^I q \;, \quad (a_1)_\mu^I = \bar q  \ga_\mu \ga_5T^I q \;,
\end{equation}
while the $B$-meson $J^{PC} = 0^{-+}$ can be interpolated by $J_B = \bar b \ga_5 q$. 
The correlation function\footnote{Our notation follows Minkowski space conventions, and, since our argument is formal, we shall ignore  questions of convergence of the path integral. 
The argument would also apply to Euclidean space, in so far as one manages to obtain the 
\ac{LD} matrix elements from this formalism. Finite-width effects of the vector mesons are not crucial for the argument, and partly cancel in the symmetry limit.}
\begin{equation}
\label{eq:M}
{\cal M}^{[V]}_{(v,a)}
\equiv  \matel{0}{T J_B(x) V^I_\mu(y) h_{\eff}(0)}{0 } \;, 
\end{equation}
provides all the necessary information to understand the properties of the matrix element 
$\matel{V}{H_{\eff}}{B}$, e.g. by using the LSZ formalism or dispersion representations in case finite-width effects are to be studied. 
Above, $V^I_\mu$ stands for either of the interpolating currents defined in \eqref{eq:inter}, 
and  
\begin{equation}
\label{eq:heff}
h_{\eff} = \bar q ( v + a \ga_5) \Gamma b \, O_r  \;,
\end{equation}
 is a schematic substitute for $H_{\eff}$ \eqref{eq:Heff}, where
$(v,a)  =  ( 1 ,\pm 1)$ correspond to $O$-  and $O'$-like operators. 
 \ac{SD} and \ac{LD} charm contributions correspond to $O_r \sim 1 , \bar c \Gamma_r c$ for example. 
 Lorentz/colour 
contractions over $\Gamma$ and $\Gamma_r$  are suppressed, as these have no impact on our argument.
Integrating out the quarks in the path integral, the matrix element \eqref{eq:M} assumes the form 
\begin{equation}
\label{eq:p-integral}
{\cal M}^{[\rho^0]}_{(v,a)} \sim  \int D \mu_\GG \Tr[ (v+a \ga_5) S_G^{(b)}(0,x) \ga_5 S_\GG^{(d)}(x,y) \ga_\mu S_\GG^{(d)}(y,0)] \;,
\end{equation} 
where $D \mu_\GG = D \GG_\mu \det( \slashed{D} + i M_f) e^{i S(\GG)} $ ($ D_\mu = (\partial - ig G)_\mu$) is 
the path integral measure, $ S(\GG)$ the Yang-Mills action, and $M_f$ denotes the mass matrix, which comprises of all flavours.  
  The quark propagator in the gluon background field is
  $S^{(q)}_\GG(w,z) = \matel{w}{(\slashed{D}  +i m_q )^{-1}} {z}$,
  and obeys 
  \begin{equation}
\label{eq:ga5}
\ga_5 S^{(q)}_\GG(w,z) = - S^{(q)}_\GG(w,z) \ga_5 \;,
\end{equation}
in the restoration limit \eqref{eq:restore}. We stress that \eqref{eq:ga5}, upon which 
the argument is based, necessitates both the vanishing $\SU(2)_A\times \U(1)_A$-violating condensates and the limit $m_q \to 0$, with some more detail on related matters deferred to 
appendix \ref{app:pi}.

\subsubsection{Relating matrix elements of parity doublers} 

Now comes the main step, where we replace $ \ga_\mu \to  \ga_\mu (\ga_5)^2$ and use
\eqref{eq:ga5} to arrive at  
\begin{equation}
\label{eq:Mav}
 {\cal M}^{[a_1]}_{(a,v)}  = -  {\cal M}^{[\rho^0]}_{(v,a)} \;,
\end{equation}
which can also be written symbolically as  
$H_{\eff}^{V \pm A   }|_{\rho}  \to  \pm H_{\eff}^{V \pm A   }|_{a_1}$. 
This relation translates into the amplitudes \eqref{eq:amp-pol}, 
with $C O$- and $C' O'$-contributions in  \eqref{eq:Heff}:
\begin{equation}
\label{eq:Ampav}
\bar{\amp}^{\bar{B} \to \rho \ga}_\chi(C,C') = \bar{\amp}^{\bar{B} \to a_1 \ga}_\chi(-C,C') \;.
\end{equation}
Moreover, in terms of the breakdown \eqref{eq:tab-break1}, the relations \eqref{eq:Mav} and \eqref{eq:Ampav} lead to
\begin{equation}
\label{eq:tab-break2}
 \begin{tabular}{ l  |  l  l   |  l  l  }
$\bar{ \amp}^{\bar{B} \to \rho (a_1) \ga}_\chi$    & $\bar{ \ampt}^{\VmA} _{SD,\chi} $         
  & $ \bar{ \ampt}^{\VmA} _{LD,\chi}$     &   $\bar{\ampt}^{\VpA} _{SD,\chi}$ & 
  $\bar{ \ampt}^{\VpA} _{LD,\chi} $ \\[0.2cm]
  \hline 
$\chi = L$ & $\pm 1$  &  $ \pm \laCKMn{i}\epsQCD{i}_{V,L}$  & $0$  & $\laCKMn{i}\epsQCDp{i}_{V,L}$ \\
$\chi = R$  & $\phantom{\pm} 0$ &  $\pm \laCKMn{i}\epsQCD{i}_{V,R}$  & $\hat{m}_{d,s} + \DRHC $ 
&  $\laCKMn{i}\epsQCDp{i}_{V,R}$  
\end{tabular}  \;.
\end{equation} 
We wish to emphasise that the procedure in this section leading to \eqref{eq:Mav} works for any local operator of the form in \eqref{eq:heff}, and so, in particular, applies to both the \ac{LD} and \ac{SD} contributions, which is reflected
in the second amplitude breakdown \eqref{eq:tab-break2}. This argument establishes the main point 
of this work, and we now turn to the discussion of how this can be applied beyond the symmetry limit.

\section{Beyond the symmetry limit}
\label{sec:beyond}

We briefly discuss the question of how to isolate the relevant phenomenology using parity doubling.\footnote{See also appendix \ref{app:PD} for more details on parity doubling.} 
\subsection{Parity doubling for phenomenologically relevant vector mesons}
\label{sec:doubleQCD}

Table \ref{tab:doublers} indicates the main $B \to V(1^-)  $ and $J^P = 1^+$ $B \to A(1^+)  $ final states, 
which we refer to as parity doublers in this work.   Note that the charge quantum number is not of importance 
for practical purposes, so that $1^{++}$ and $1^{+-}$ 
 can both be considered as effective parity doublers of the $1^{--}$ (or $1^{-+}$, cf. caption of table \ref{tab:doublers}) states. 
The interpolating operators,  denoted $O_V$ in the table, are 
\begin{equation}
\label{eq:inter2}
V_{(5)}^{(I)} = \bar q \ga_\la  (\ga_5) (T^I) q  \;, \quad T_{(5)}^{(I)} = \bar q \sigma_{ \kappa \la}( \ga_5) (T^I) q  \;,
\end{equation}
where $q=u,d$, and Lorentz indices have been suppressed on the left-hand side.
Vector and axial mesons couple to these currents as 
\begin{equation}
\label{eq:current}
\matel{0}{V_{(5)}}{V(A)} \sim \eta_\la \;, \quad \matel{0}{T_{(5)}}{V(A)} \sim  \eta_{[\la} p_{\kappa]} \;,
\end{equation}
where square brackets denote antisymmetrisation in indices.   
The canonical parity doublers are listed on the horizontal lines. 
However, by $\SU(N_F)_V$ flavour symmetry, it will become clear that the measurement of a single parity doubler 
can reveal important information on primed Wilson coefficients $C'_{7,8}$ (and $C'_{9,10}$) by 
disentangling the   \ac{LD} contributions.

\subsection{Sources of correction to the symmetry limit}
\label{sec:ratios}

In the real world, $\SU(N_F)_A\times \U(1)_A$ is broken, e.g. by
$ \vev{\bar q q }_{\mu = 1 { \textrm{\footnotesize GeV}} } \simeq (-0.24(1) \GeV)^3$, 
raising the question of the corrections  to~\eqref{eq:Mav} and the resultant breakdown \eqref{eq:tab-break2}.
We find it advantageous to distinguish two sources: corrections to the $\ga_5$-trick \eqref{eq:ga5}, and those 
arising from the hadronic parameters of the vector mesons.

Corrections to \eqref{eq:Mav}, for instance, can be understood by considering the
light-cone \ac{OPE}, with the interpolating current replaced by a $B$-meson light-cone \ac{DA} e.g.
\cite{Khodjamirian:2005ea}, in place of the complete path integral \eqref{eq:p-integral}. In this case, \eqref{eq:ga5} results in both $m_q$- and $\vev{\bar q q}$-corrections, along with other condensates. Whereas the former
are parametrically small, the latter are  suppressed in effect by ${\cal O}(1)| \vev{\bar q q} /M^3 | \simeq 
{\cal O}(1)(1/4)^3$, where $M \simeq 1 \GeV$ is the Borel mass  scale. 
It remains to be seen how effective this  suppression is  in explicit computations.
The remaining corrections to the symmetric breakdown  \eqref{eq:tab-break2} arise from differences in 
the hadronic parameters, namely the meson mass and the  \ac{DA} parameters. 
The latter are either (partly) known or are expressible in terms 
of $m_q$, $\vev{\bar q q}$, and higher-dimensional condensates \cite{prep_da}.  Moreover, these can be assessed experimentally,  
cf. section \ref{sec:app}.

\begin{table}[t]
\centering
 \begin{tabular}{l | l   l  l || l  |  l   l l | l |  l  l l    }
$ I^G $ 
& $1^{--}$   & $\frac{\Gamma_V}{m_V}$  & $O_V$ &
$I^G$ 
&  $1^{++}$   &$\frac{\Gamma_V}{m_V}$  & $O_V$ &
$I^G$
& $1^{+-}$  & $\frac{\Gamma_V}{m_V}$ & $O_V$ 
   \\[0.1cm] \hline
$1^+$   
&   $\rho(\text{\small{770}})$        & $19.1(1)$ & $(V,T)^I$   &
$1^-$  
& $a_1(\text{\small{1260}})$  & $35(14)$ & $V^I_5$ &
$1^+$  
& $b_1(\text{\small{1235}})$   & $11.5(7)$ & $T^I_5$  \\[0.1cm] \hline
$0^-$   
 & $\omega(\text{\small{782}})$  &  $1.08(1) $ &  $V,T$ & 
 $0^+$
 & $\foneone$   &  $1.77(1)$       &  $V_5$ &
 $0^-$
 & $\honeone$  &  $31.0(5)$ &  $T_5$ \\
 $0^-$   
 & $\phi(\text{\small{1020}})$     &  $0.417(2)$ & $(V,T)^{\bar ss}$ &
 $0^+$
 & $\fonetwo$  &  $3.8(2) $&    $V^{\bar s s}_5$ &
 $0^-$
 & $\honetwo$  & $6.3(16)$ &   $T^{\bar s s}_5$                \\[0.1cm] \hline
$I$  &  $1^{-}$  &       &   &       & $1^{+}$   &   & & &  $1^{+}$  &    \\[0.1cm] \hline
  $\frac{1}{2}$   
 &   $K^*(\text{\small{895}})$      & $5.6(1)$    & $(V,T)^{s}$  &
  $\frac{1}{2}$ 
 &   $\Koneone$    & $7.1(16)$  & $V^{s}_5$  &
  $\frac{1}{2}$ 
 &   $\Konetwo$   & $12.0(9)$    & $T^{s}_5$ 
 \end{tabular} 
\caption{\small Mass ($m_V$) and width ($\Gamma_V$) data for the neutral mesons, from the latest \acs{PDG} data \cite{PDG2016}. 
Uncertainties in these parameters are also indicated in brackets alongside the central values.
The $I^G$ and $J^{PC}$ quantum numbers have also been indicated.  
The $K$ particles are separated, as they do not have definite $G$-parity states. 
The two $K_1$-particles are subject to a mixing angle $\theta_{K_1}$, which is fortunately  
known to reasonable accuracy \cite{Cheng2011}. Mixing between particles in the same column has to 
be taken into account as well, cf.  appendix C of \cite{BSZ2015} for the discussion of the $1^{--}$-states for example.
 The interpolating operators under the $O_V$-column are described in the main text 
in \eqref{eq:inter2} and superscripts  $s$ and $\bar s s$  denote replacements of the light quarks $q \to s$. 
In addition, there are ``exotic'' $1^{-+}$ states, with interpolating operators of the type  
$O_V \sim  \bar q \ga_\la   T^I  \Dlra_{\mu}  q$ and $\bar q \sigma_{\la \kappa}T^I  \Dlra_{\mu} q$. The 
$\pi_1(\text{\small{1400}})$ \cite{PDG2016} $(I^G = 1^-)$ 
 is a candidate particle for carrying these quantum numbers. Such states are rather broad, e.g. $\Gamma_{\pi_1}/m_{\pi_1} \sim 1/4$, and are not well-studied. 
 The particle content of some non-exotic states in the table remains to be definitively established \cite{PDG2016}, but all states are expected 
 to have significant $\bar q q$-wave functions. 
 For example, from ALEPH data \cite{Schael:2005am}, e.g. figures 63 and 65, it can be inferred that 
 the statement above is correct for $a_1(1260)$.  
 The determination of the $3$-particle content, or the matrix element $\langle 0 | \bar q G q |V(A) \rangle$, is an 
 important problem which deserves assessment from lattice QCD, in addition to the existing QCD sum rule results.}
\label{tab:doublers}
\end{table}

\section{Applications to experimental searches} 
\label{sec:app}

\subsection{Right-handed currents from time-dependent rates} 
\label{sec:TDCP}

A relevant question is how to test the chirality hierarchy \eqref{eq:tab-break1}. 
The decay rate  does not lend itself to such tests, as 
the right-handed amplitude is dominated by the left-handed one. 
The situation is, however, favourable in the case where 
the $B$-meson is neutral and undergoes mixing \cite{AGS97} (and/or decays to at least 3 hadrons
and a photon \cite{Gronau:2001ng,Atwood:2004jj,Becirevic:2012dx,Gronau:2017kyq}). The mixing  is
driven by particle and antiparticle having a common final state, e.g.
 $\bar{B} \to V\ga_L  \leftarrow  B$. In this case, one of the amplitudes 
 is chirally suppressed, giving rise to a direct  
  linear behaviour in right-handed amplitudes,  $\Gamma_{\textrm{mix}} \sim \bar{\amp}_R$, 
  compared to the unfavourable behaviour of the $t=0$-rate 
  $\Gamma \sim |\bar{\amp}_L|^2 +  |\bar{\amp}_R|^2$. 
 
The time-dependent rate of a $B_\DD$ meson, produced at $t=0$, 
assuming CPT-invariance and $|q/p|=1$,\footnote{The quantities $p$ and $q$ 
describe the transition matrix from mass to  flavour eigenstates 
where $\textrm{arg}(q/p) = - \phi_{B_D}$.  In $B_\DD$-$\bar{B}_\DD$ mixing they are indeed compatible with the assumption $|q/p|=1$, up to negligible corrections \cite{HFAG2016}.}
takes the form
\begin{equation}
\label{eq:bbar-t}
{\cal B}(\bar{B}_\DD[B_\DD] \to V \ga) = 
     B_0 e^{-\Gamma_\DD t}[{\textrm{ch}}(\frac{\Delta \Gamma_\DD}{2}t) -\HH { \textrm{sh}}(\frac{\Delta \Gamma_D}{2}t)
                \mp C\cos(\Delta m_{\DD} t) \pm S \sin(\Delta m_\DD t)] \;,
\end{equation}
where $\Delta \Gamma_\DD \equiv \Gamma^{(H)}_\DD- \Gamma^{(L)}_\DD $ is the width difference, and $\Delta m_\DD \equiv m^{(H)}_\DD- m^{(L)}_\DD $ the mass difference, of the heavy ($H$) and light ($L$)  mass eigenstates.  
The quantities $S$ and $C$ are related  to indirect and direct CP violation respectively. In the \ac{PDG} notation \cite{PDG2016}, $H \equiv {\cal A}^{\Delta \Gamma}$.
In terms of the decomposition \eqref{eq:amp-pol}, dropping the superscripts for brevity, these quantities read
(${\cal N} =  |\amp_L|^2 + |\bar \amp_L|^2 
+  |\amp_R|^2 + |\bar \amp_R|^2$)
\begin{alignat}{2}
\label{eq:SHdef}
& S(H)  &\;=\;&   2   { \Ima(\Rea) }\left [\frac{q}{p}(\bar \amp_L \amp_L^* +
\bar \amp_R \amp_R^*) \right]  {\cal N}^{-1}   \;,  
 \end{alignat}
and we quote
$C =  \left( (|\amp_L|^2 + | \amp_R|^2)
-  (|\bar \amp_L|^2 +  |\bar \amp_R|^2 ) \right) {\cal N}^{-1}  $, although this observable is of no further relevance for this work.

In the \ac{SM}, there are three weak phases orginating from $\la_{u,c,t}$, one of which 
can be eliminated by the unitarity relation $\la_u + \la_c + \la_t = 0$. 
Hence, one may write
\begin{alignat}{6}
\label{eq:ALR}
& \bar{\amp}_L &\; \sim\;&  (1 &\;+\;&  \laCKMn{i} \, \epsQCD{i}_{V,L} )  \quad &\Rightarrow & \quad
 \amp_R \sim \xi_{V} (1 &\;+\;& \laCKMn{i}^* \,\epsQCD{i}_{V,L} )\;, \nonumber \\[0.1cm]
& \bar{\amp}_R &\; \sim\;&  (\hat{C_7'} &\;+\;&  \laCKMn{i} \, \epsQCD{i}_{V,R} )  \quad &\Rightarrow & \quad
 \amp_L \sim \xi_{V} (\hat{C_7'} &\;+\;&  \laCKMn{i}^* \,\epsQCD{i}_{V,R} )\;,
\end{alignat}
where the result on the right follows by CP conjugation, and $\xi_{V}$  is the  CP eigenvalue of $V$.   
 The observables $S$ and $H$ take the form
 \begin{alignat}{3}
 \label{eq:SHmaster}
& S(\HH)_{V(A)\ga} &\;=\;&  2 \xi_{V} \{ &\pm & ( \hat{m_\DD} \sincos (  2 \phi_t - \phi_{B_\DD})  +
 \Delta_R \sincos (  2 \phi_t +  \phi_R- \phi_{B_\DD} )) 
 \; +   \nonumber \\
&   & &   & & 
| \tilde{\la}_i| {\Rea}[ \epsQCD{i}_{V(A),R} ]  \sincos (   \phi_t + \phi_i- \phi_{B_\DD} ) \}
(1   + {\cal O}(\hat{m}_D,\Delta_R, \epsQCD{i}_{V(A),\chi} ))  
  \;, 
\end{alignat}
where the sines and cosines refer to $S$ and $H$, and the signs $\pm$ follow from 
the breakdown \eqref{eq:tab-break2}. Corrections to \eqref{eq:SHmaster} can be expected to be small, and are not difficult to restore, but we have chosen to present a simple formula for illustrative purposes.

It is instructive to expand \eqref{eq:SHmaster} for specific modes. There are four classes of $B \to V \ga$ 
decays, due to the choice of initial state meson, $B_d$ or $B_s$, while  
the transition itself can be either $b \to d $ or $b \to s$.  Since $\Delta \Gamma_d$ is too small 
to have an observable effect, this makes the $\HH_{B_d}$ parameter unobservable 
in practice. Below, we present the observables \eqref{eq:SHmaster} for three of these classes, postponing details on  
$B_s \to \bar{K}^*$ ($b \to s$) to \cite{prep_charm} as the decay is experimentally less attractive. 
The formulae in \eqref{eq:SHmaster} take the form
\begin{alignat}{2}
\label{eq:SHex}
& S_{B_d \to \rho(a_1) \ga} &\;\simeq\;&   2 \{  \pm \Delta_R \sin \phi_R
 +  \sin \be |\tilde{\la}_c^{(d)}| \Rea [\epsQCD{c}_{\rho(a_1),R}] -
\sin (\be + \ga)  |\tilde{\la}_u^{(d)}| \Rea [ \epsQCD{u}_{\rho(a_1),R}]     \} \, , \nonumber \\[0.2cm]
& S_{B_d \to K^*(K_1)\ga} &\;\simeq\;&  2 \xi_{K^*(K_1)} \{  \mp (   
 \Delta_R \sin (2 \be - \phi_R)+  \hat{m}_s \sin 2 \be ) + \sin 2\be \, \Rea [ \epsQCD{c}_{K^*(K_1),R}]   \}  \, , \nonumber \\[0.2cm]
& S_{B_s \to \phi(f_1)\ga}  &\;\simeq\;&     2  \{ \pm \Delta_R \sin ( \phi_R)  \}   \, , \nonumber \\[0.2cm]
& \HH_{B_s \to \phi(f_1)\ga}  &\;\simeq\;&    2  \{   \pm (\Delta_R \cos ( \phi_R) +  \hat{m}_s  ) - \Rea [\epsQCD{c}_{\phi(f_1),R} ] \} \,,  
\end{alignat}
where 
$ \tilde{\la}_{c,u}^{(d)} = O(1) $, $ \tilde{\la}^{(s)}_u \ll \tilde{\la}^{(s)}_c \equiv \la_c^{(s)}/ \la_t^{(s)} \simeq - 1$, 
and we have used $\xi_{V,A}  = 1$. For $S_{B_d \to K^*(K_1)\ga}$, either (near) CP eigenstate 
($K_{S,L} \pi^0$)
can be observed in the subsequent decay, so we indicate the explicit CP eigenvalue $\xi_{K^*(K_1)}$ in this case.\footnote{Taking the difference in  $S_{B_d \to K^*(K_S\pi^0)} - S_{B_d \to K^*(K_L\pi^0)}$ 
will enhance the statistics, but cannot eliminate the \ac{LD} contribution, as the CP-eigenvalue is just a global phase.}
The vanishing of $S_{B_s \to \phi(f_1)\ga} \simeq 0$ in the \ac{SM}
comes from the cancellation of all weak phases involved, and this quantity is therefore a null test for weak 
phases of \ac{RHC}.

The expressions \eqref{eq:SHex} are for the $J^{PC} = 1^{++}$ parity doublers. If, instead, one were to use doublers with the $J^{PC} = 1^{+-}$ quantum numbers, then $\xi_{(1^{+-})} = -1$, and the corresponding observables $S$ and $H$ pick up an additional minus sign.

\subsection{The approach exemplified in $B_s \to \phi(f_1)\ga$}
\label{sec:exemplified}

From \eqref{eq:SHex}, it can be seen that both observables $S$ and $H$ in \eqref{eq:SHmaster} are of interest, and can be observed experimentally, in the decay $B_s \to \phi(f_1/h_1)\ga$, where $f_1 \equiv \fonetwo$ and $h_1 \equiv \honetwo$. 
Using $\xi_{h_1} = -1$, one gets a remarkable equation 
\begin{alignat}{2}
\label{eq:Hcharm1}
& \left(  \HH_{\phi\ga} \pm  \HH_{\fone(\hone)\ga} \right) &\;\simeq\;& 
- 2 {\textrm{Re}}[ \epsQCD{c}_{\phi,R} +  \epsQCD{c}_{\fone(\hone),R}   ]  \nonumber \\[0.1cm]
& &\;=\;& 
- 2 {\textrm{Re}}[ \epsQCD{c}_{\phi,R}] ( 1 +  \RR_{\fone(\hone),\phi}^c    )  \;,
\end{alignat}
where
\begin{equation}
\label{eq:RR}
\RR_{A,V}^i \equiv \frac{\textrm{Re} [\epsQCD{i}_{A,R}]}{\textrm{Re} [ \epsQCD{i}_{V,R}]}  
= 1 + {\cal O}(m_q,\vev{\bar q q}) \;.
\end{equation}
and the $\simeq$ in \eqref{eq:Hcharm1} indicates the approximations made in \eqref{eq:SHmaster}, as well as neglecting the 
$\la_u$ LD contribution.\footnote{These approximations can be easily relaxed if required, depending on the required accuracy, cf. also the discussion following Eq.~\eqref{eq:SHmaster}.}
Eq. \eqref{eq:Hcharm1} is remarkable in that it shows that 
it is possible to measure the sum of the \ac{LD} (charm) contributions 
without any compromise from the symmetry breaking, thanks to the exact form factor relation $T_1(0) = T_2(0)$.  In extracting the \ac{LD} contribution, what matters is not how far away $\RR_{A,V}$ is from unity but 
the uncertainty itself. As an example, supposing we could determine $\RR_{A,V}$ to $20\%$ uncertainty, with 
 $\RR_{A,V} $ being one of the four values $(1,1.2,1.5,2)$. Then one could extract $\epsQCD{c}_{\phi,R}$ from experiment with an accuracy 
 of $(10,11,12,13)\%$ respectively. This is very much improved situation in two ways. 
 Firstly, one can extract the \ac{LD} contribution by solely predicting the uncertainty on $\RR_{A,V}^i$ rather than 
 the \ac{LD} matrix element itself. In addition, this translates into a considerably smaller uncertainty of the
 \ac{LD} matrix element than one could  hope to get from an a priori computation.

Finally, we mention that the counterpart of \eqref{eq:Hcharm1} is the equation where the 
\ac{SD} part is enhanced and the \ac{LD} part reduced, and reads
\begin{alignat}{2}
\label{eq:HRHC1}
 \Delta_R  \cos(\phi_{\Delta_R}) =  \frac{1}{4} ( 
    \HH_{\phi\ga} \mp  \HH_{\fone(\hone)\ga} )  + 
\frac{1}{2} \Rea [ \epsQCD{c}_{\phi,R}  - \epsQCD{c}_{\fone(\hone),R}]  
 - \hat{ m}_s
    \;.
\end{alignat}
It remains to be seen in the future how well such quantities can be measured and how well
the ratios $\RR_{A,V}^i$  can be predicted.  It is clear that the potential improvement is significantly 
larger than what can be hoped for in a direct computation, as many uncertainties will cancel, some of 
them due to the symmetry limit.

 \subsection{$B \to V \ell \bar{ \ell}$ and other decay channels}
 \label{sec:other}
 
Decays such as $B \to K^*(\to K \pi) \mu^+ \mu^-$ are  an important probe of \ac{NP} in the flavour sector \cite{LHCbKstar2015,Belle2016Dec,CMSBtoKs2018,ATLASBtoKs2018}, as their angular distributions allow the assessment of 
a wealth of observables, twelve for the dimension-six $H_\eff$ \cite{KSSS99}; higher-dimensional 
operators are also accessible by extracting higher moments (modulo QED effects) \cite{GHZ2016}. 

The parity-doubling approach can be extended to $B \to (V,A) \ell \bar{\ell}$
rather straightforwardly by considering the angular observables.  In the notation of \cite{GHZ2016}, 
the moments, or angular coefficients, are $\mathbb{G}_m^{l_k,l_\ell}$, where $l_{k,\ell}$ denote the partial 
wave of the $K\pi$ and $\mu\mu$-pair respectively, and $m$ is the relative helicity difference. An interference 
of left- and right-handed polarisation corresponds to the helicity difference $m =2$, which leads to the real and imaginary parts of $\mathbb{G}_m^{l_K,l_\ell} \to \mathbb{G}_2^{2,2} $ being the observables of interest. These have indeed been identified, with the acronyms 
$P_1 = A_T^{(2)} \sim \Rea[\mathbb{G}_2^{2,2}]$,  
$P_3 \sim  \Ima[\mathbb{G}_2^{2,2}]$, some time ago (e.g.  \cite{KM05,Becirevic:2011bp})  
as giving access to \ac{RHC} at low $q^2$.\footnote{The restriction to low $q^2$ is due to kinematics and matrix element effects. 
Firstly, the kinematics of Lorentz invariance enforces that the helicity amplitudes are degenerate at maximal $q^2$ (the kinematic endpoint) \cite{HZ2013}, so that the V-A effect is maximally diluted. Secondly, the accidental control $T_1(0) = T_2(0)$ is weakened
when $q^2$ is increased, as the \ac{LD} effect enters the left-handed amplitude.} 
A measurement of the right-handed \ac{LD} contribution at $q^2 =0$, or at low $q^2$, could also 
provide invaluable information on the angular anomalies \cite{round} (most notably, $P_5' \sim  \Rea[\mathbb{G}_1^{2,1}]$) observed in $B \to K^* \ell \bar{\ell}$ at LHCb, in connection with approaches using analyticity 
\cite{LZ2014,Bobeth:2017vxj}.  
In this respect, $B \to K^* e^+ e^-$ is an even more promising channel, studied at the 
LHCb experiment  \cite{LHCbBKee2015}, and with good prospects at Belle II. 
 Exploring the potential of time-dependent angular distributions
  would also seem to be an interesting possibility \cite{DV15}.
 
Thus, $B \to (V,A) \ell \bar{\ell}$ allows the assessment of all the right-handed operators $O'_{7,8,9,10}$, 
or their respective Wilson coefficients, without resorting  to time-dependent amplitudes, meaning that decay modes for charged mesons can also be assessed in this manner.  

We wish to emphasise that the ideas in this paper, whilst they have been discussed primarily in the context of $B \to (V,A) \ga$ and $B \to (V,A) \ell \bar{\ell}$ decays, can be extended to other decays of interest, as long as such systems admit parity-doubling partner decays.\footnote{This excludes the Kaon system, for example, which decays into pions; being pseudo-Goldstone bosons, these do not have parity-doubling partners.} 
Examples include the $D \to V \ga (\ell \bar \ell)$ sector, e.g. \cite{Lyon:2012fk,deBoer:2018zhz}, as well as higher-spin states e.g. $B \to K_2 \ga (\ell \bar \ell) $, since parity doubling occurs in those modes \cite{Afonin:2007mj}.\footnote{A hybrid of the two types of decay discussed above is the photon conversion 
 $B \to (K^* \to K \pi) (\ga N \to \ell^+ \ell^- N)$,  proposed in \cite{RobinsonBishara15}, for which the methods of 
 this paper apply equally when the $K^*$ is replaced by the $K_1(1270)$, for example.}  Applications to symmetry-based amplitude parametrisations in $B \to P V(A) $, 
with $P$ being a pseudoscalar meson, are another possibility, although one would expect stronger breaking 
effects in final-state interactions than in the \ac{LD} amplitude of radiative decays.

\section{Discussion and conclusions}
\label{sec:conc}

In this paper, we have advocated that the contamination of \acl{RHC} in $ B \to V \ga(\ell \bar{\ell})$ decays due to \acl{LD} effects can be controlled by considering in addition the corresponding decay $ B \to A \ga(\ell \bar{\ell})$. 
It was shown, in the chiral symmetry restoration limit \eqref{eq:restore}, that the leaking of the V-A contributions into 
the right-handed  amplitude for the parity doublers comes with exactly the opposite sign, 
 as compared to the leading \acl{SD} contributions  \eqref{eq:tab-break2}. 

The case  beyond the symmetry limit is briefly discussed in the previous section 
for $B \to V(A) \ell \bar{\ell}$, and for $B \to V(A) \ga$ the summary is as follows.
The sum of the vector and axial \acl{LD} contribution can still
be extracted from experiment (e.g. \eqref{eq:Hcharm1}),  because of the exact form factor 
relation $T_1(0) = T_2(0)$. 
This then allows to test theory predictions of exclusive \acl{LD} contributions \cite{BZ06CP,MXZ2008,Khodjamirian:2010vf,prep_charm} and contrast them with the, somewhat larger, indirect estimation from the 
inclusive $b \to X_s \ga$ channel  \cite{GGLZ2004}. More concretely, the charm contamination was computed  
to be $0.6\%$ for $S_{B_d \to K^* \ga} $ in \cite{BZ06CP}, whereas \cite{GGLZ2004} 
estimated the contaminations to be up to $6\%$. In addition, the smallness of the theoretical result 
can be understood by invoking  the parity-doubling limit  \cite{prep_charm}. 
 
From \eqref{eq:Hcharm1} one can extract the sum of the axial and vector \acl{LD} contribution. 
One can obtain the individual contributions by taking into account a theory prediction of the 
ratio $\RR_{A,V}$ \eqref{eq:RR}, which comes with reduced uncertainty due to standard uncertainty cancellations, further enhanced by the symmetry limit. An error on $\RR_{A,V}$ of $20\%$ results in an uncertainty on the single
\acl{LD} contribution of just around $10\%$, if no experimental error is assumed. In summary, we have proposed 
a data-theory-driven program to reduce the uncertainty of the \acl{LD} contamination, paving the way 
to much cleaner searches for \acl{RHC}.\footnote{In reference \cite{Atwood:2004jj} it was suggested, 
though not worked out in detail, that the use of $B \to P_1 P_2 \ga$ together with a Dalitz-plot analysis could 
be used to disentangle the \acl{LD}- from the \acl{SD} contribution.}

Let us turn at last to the experimental perspectives.
For Belle II, an uncertainty of $3\%$ is anticipated in this observable with a data set of 
$50 \, \textrm{ab}^{-1}$  \cite{Urquijo}. 
This estimate does not yet take into account the gain in photon efficiency in differences and sums of rates, 
relevant to the parity-doubling approach cf. (\ref{eq:Hcharm1},\ref{eq:HRHC1}).
The analysis of axial vector meson 
final states is more challenging, because of their decay chains. However,  the measurement 
of a single channel can still provide invaluable information on the \acl{LD} contributions \eqref{eq:Hcharm1}, 
as they are related by $\SU(N_F)_V$ flavour symmetry, taking into account normalisation issues such as 
mixing angles.
Promising  states are the $K_1(1270)$ and the $\foneone$ or $\fonetwo$.
Belle has reported a time-dependent measurement of  $B^0 \to \rho^0 K_s \ga$ \cite{Li:2008qma} and
the rate  ${\cal B} (B^+ \to K^+_1(1270) \ga) \simeq  4.3(9)(9) \cdot10^{-5}$  
\cite{Yang:2004as}. At the LHCb experiment, the
 $K_1$ states were seen in $B \to K^+ \pi^+ \pi^-$ at the LHCb \cite{Aaij:2014kwa},
and the first observation of ${\cal B}\left ( B_{s,d} \to  \foneone J/\Psi \right) \simeq 7 \cdot 10^{-5}, \, 8\cdot10^{-6}$ was reported in \cite{Aaij:2013rja}.

\subsection*{Acknowledgements}

JG and RZ are grateful to Gino Isidori for an extended and refreshing stay at the Pauli-Centre for Theoretical Physics 
at the University of Z\"urich, where a large fraction of this work has been undertaken. 
  RZ would like to thank Johannes Albrecht, Greig Cowan, Akimasa Ishikawa, Franz Muheim,  Christopher Smith,  and
Philip Urquijo,
as well as many of the participants of  QCD-Moriond-2018 and  the
$b \to s \ell \ell$ workshop at Miapp (TU Munich), for useful discussions.
JG acknowledges the support of an STFC studentship (grant reference ST/K501980/1).

\appendix

\section{Chiral restoration limit and  $ \ga_5 S_G^{(q)}= - S_G^{(q)} \ga_5$ \eqref{eq:ga5}}
\label{app:pi}

The pion decay constant, $F_\pi \simeq 92 \MeV $ in QCD, 
is defined by $ \matel{\pi^b(p)}{ J^{A,a}_\mu }{0}    = \de^{ab}  p_\mu  F_\pi$, 
with $ J^{A,a}_\mu = \bar q T^a \ga_\mu \ga_5 q$, and  $q$ are  $N_F$ light  quarks.  
$F_\pi $ is  the order parameter of spontaneous chiral symmetry breaking. 
At the heuristic level, $F_\pi \neq 0$ implies the non-invariance of the  vacuum 
with respect to the  axial flavour charge $Q^{A,a} = \int_V d^3 x J^{A,a}_0 $. 
In this appendix, we aim to show that our formula \eqref{eq:ga5},
\begin{equation}
\label{eq:iff}
m_q , \vev{\bar q q} = \dots = 0  \quad  \Leftrightarrow \quad  \ga_5 S_G^{(q)}(w,z) = - S_G^{(q)}(w,z) \ga_5 \quad \left( \Leftrightarrow \quad  F_\pi =0  \right) \,, 
\end{equation}
depends, as expected, on the restoration limit.  This result serves to illustrate 
\eqref{eq:ga5} in some more detail than discussed in the main text.

We first proceed to show that $\ga_5 S_G^{(q)}(w,z) = - S_G^{(q)}(w,z) \ga_5 \quad  \Rightarrow \quad  F_\pi =0$. 
For this purpose, consider the correlation function, used to derive the Weinberg sum rules  \cite{WSR},
\begin{equation}
\label{eq:start}
(\Pi^{a,b}_{LR})_{\mu \nu}(q^2)  = i \int d^4 x e^{i q \cdot x} \vev{T J^{a,L}_\mu(x) J^{b,R}_\nu(x)}  = (q_\mu q_\nu - q^2 g_{\mu \nu} )
\Pi^{a,b}_{LR}(q^2) \;,
\end{equation}
where $J^{a,L(R)}_\mu =  2 \bar q T^a \ga_\mu (\ga_5) q_{L(R)}$, and $(a,b)$ are $\SU(N_F)_{(L,R)}$ flavour indices.  
 One may integrate out the fermion in the path integral formulation 
to obtain 
\begin{equation}
\label{eq:VA}
(\hat{\Pi}^{a,b}_{\textrm{LR}})_{\mu\nu}(x) =  \frac{1}{2} \de^{ab}  \int D \mu_\GG
\left(  \textrm{Tr}[S_G^{(q)}(x,0) \ga_\mu  
S^{(q)}_G(0,x) \ga_\nu ] - 
 \textrm{Tr}[S^{(q)}_G(x,0) \ga_\mu \ga_5  
S^{(q)}_G(0,x) \ga_\nu  \ga_5]   \right) \;, 
\end{equation}
where the hat denotes the Fourier transform, and  the path integral measure, already defined below Eq.~\eqref{eq:p-integral}, is $D \mu_\GG = D \GG_\mu \det( \slashed{D} + i M_f) e^{i S(\GG)} $. 
Using  $\ga_5 S_G^{(q)}(w,z) = - S_G^{(q)}(w,z) \ga_5$ \eqref{eq:ga5},  one concludes 
that  $(\hat{\Pi}^{a,b}_{\textrm{LR}})_{\mu\nu}(x)  = 0$ by commuting 
the $\ga_5$ through the fermion propagator and the $\ga$-matrix in \eqref{eq:VA}.
Since $4 (\Pi^{a,b}_{LR})_{\mu \nu} = 
\de^{ab} ( (\Pi_{VV})_{\mu \nu} - (\Pi_{AA})_{\mu \nu})$, 
the spectral densities  
\begin{equation}
\label{eq:V=A}
\rho_V(s) = \rho_A(s) \;, 
\end{equation}
 of the dispersion representation ($\pi \rho_{V,A}(s) = \Ima[ \Pi_{VV(AA)}(s)] $)\footnote{
Strictly speaking, one would need to subtract the dispersion representation once, but it is immaterial 
for the presentation.}
\begin{equation}
\Pi_{V(A)}(q^2) =  \int_0^\infty ds \frac{\rho_{V(A)}(s)}{s-q^2-i0} \;, 
\end{equation}
are point-by-point identical.
The spectral densities are given by 
\begin{equation}
\label{eq:rho_app}
\rho_A(s) = F_\pi^2 \de( s- m_\pi^2) + F_{a_1}^2 \de( s- m_{a_1}^2) +\dots \;, \quad 
\rho_V(s) =   F_{\rho}^2 \de( s- m_{\rho}^2) + \dots \;,
\end{equation}
where the dots stand for higher states, and the fact that we used the narrow width approximation 
for the $\rho$ and $a_1$ meson is immaterial. The only crucial point is that \eqref{eq:V=A} necessarily 
implies $F_\pi = 0$, since there is no massless particle in the vector channel.

 In the second part, we show that 
 $m_q , \vev{\bar q q} =  \CondQGQ{q}   = \dots = 0  \quad  \Leftrightarrow \ga_5 S_G^{(q)}(w,z) = - S_G^{(q)}(w,z) \ga_5$, which is what 
 was used in the main text in section  \ref{sec:path}. In practical computations, using the \ac{OPE} one 
 uses the formula \cite{PT84}
 \begin{align}
\label{eq:twoquarkConddef}
 \vev{ \bar{q}_m(x_1) q_n(x_2)} =  N_c &\left[\frac{1}{12}\left(1+ \frac{i}{d} m_q \slashed{x}_{12}\right)_{nm} \CondQQ{q}  + \frac{g^2 }{288}  \left(\frac{i}{12}x_{12}^2  (\slashed{x}_{12})_{nm} \Condqqff{q} \right) \right.  \nonumber \\[0.1cm]
& \left. +  \frac{1}{192}x_{12}^2 (1 + \frac{i}{6} m_q \slashed{x}_{12} )_{nm}  \CondQGQ{q}  + \dots \right]
\end{align}
where $x_{12} \equiv (x_1-x_2)$, $m$ and $n$ are Dirac indices, 
$\Condqqff{q} = \vev{\bar{q} \ga_\mu t^a q \sum_f \bar{f} \ga^\mu t^a f}$,  and the dots stand for higher dimensional condensates.  We only consider the terms which do not vanish in the 
$m_q \to 0$ limit. 
It is readily seen that   $\CondQQ{q}$ and $\CondQGQ{q}$ are obstructions to \eqref{eq:ga5}, which
is expected since they are not invariant under $\SU(N_F)_A$ (and $\U(1)_A$). 
The contrary  applies to $\Condqqff{q}$. 
The statement about $\vev{\bar q q}$  can be made slightly more 
rigorous. 
  Following the argument of \cite{Banks:1979yr}, the 
 fermion propagator in the external gluon field may be written as 
 \begin{equation}
\vev{q(x) \bar q(0)}_G =  i S_G(x,0) = \sum_n \frac{\phi_n(x) \phi_n^\dagger(0)}{m_q- i \la_n} \end{equation}
where $\phi_n$ are the Dirac operator eigenmodes $i \slashed{D} \phi_n = \la_n \phi_n$.
Noting that $\ga_5 \phi_n$ is an eigenmode
of eigenvalue $- \la_n$, and that  the zero eigenmodes are suppressed by $V^{-1/2}$ in the infinite-volume limit, 
one finds 
\begin{equation}
\vev{\bar q q}  = \frac{1}{V} \int_V d^4x \vev{\bar q q}   \stackrel{V\to \infty}{\to}   
 -2 m_q \int_0^\infty d \la \frac{\rho(\la)}{m_q^2 + \la^2}  \;,
\end{equation}
where the function $\rho(\la)$ is the Dirac eigenmode 
density. In the limit $m_q \to 0$, one obtains  
the celebrated Banks-Casher relation $\vev{\bar q q} = - \textrm{sign}(m) \, \pi\rho(0)$ \cite{Banks:1979yr}.   
Therefore, one concludes that $\vev{\bar q q} =0$ ($\rho(0) = 0$) is a necessary condition 
for  $ \ga_5 S^{(q)}_\GG(w,z) = - S^{(q)}_\GG(w,z) \ga_5$ \eqref{eq:ga5} to hold. 
Similar arguments would apply to further terms in \eqref{eq:twoquarkConddef}, but are more difficult 
to render rigorous. 
Since $F_\pi =0 \Rightarrow m_q , \vev{\bar q q} = \dots = 0 $, this finally results in 
\eqref{eq:iff}.

\section{Parity doubling}
\label{app:PD}

In this section, we provide some minimal background on parity doubling, 
which has a long history in particle physics \cite{Afonin:2007mj}. Parity doubling
achieved  its modern paradigm shift with the advent of the Weinberg sum rules \cite{WSR}, 
partly described in the previous section,  and has recently been investigated on the lattice \cite{DGL14,DGL15,Rohrhofer:2017grg}.\footnote{The restoration of the axial flavour symmetries  
for excited states in the spectrum was proposed in \cite{G07_PRP} and subsequently challenged in
\cite{SV07}.}$^,$\footnote{\label{foot:lattice}  Lattice simulations at temperatures 
above the chiral phase transition have been performed \cite{overlap,DW}, where a restoration of the $\U(1)_A$-anomaly has been observed. 
The restoration of the symmetry in the spectrum, along with enhanced symmetries, has 
been found in lattice simulations with truncated eigenmodes of the 
Dirac operator \cite{DGL14,DGL15}. The motivation for this truncation is that the lowest eigenvalues 
are related to the breaking of chiral symmetry by the
 Banks-Casher relation \cite{Banks:1979yr}. The restoration of the flavour symmetry, and some of the symmetry enhancements, have been confirmed by 
the above-mentioned finite-temperature simulations \cite{Rohrhofer:2017grg}. Whereas more is to be learnt 
from this interesting topic in the future, the precise outcomes 
are not important for our practical purposes. However, if one were able to perturb in $\vev{q q}$ and $m_q$, then
the exact limit would be of interest.}

The basic idea is that a global symmetry, generated by a charge $Q$,  induces  degeneracies in the spectrum, as it commutes with the Hamiltonian.
Examples include supersymmetry, with degeneracies between bosons and fermions, or simply
the global $\SU(N_f)_V$ flavour symmetry, leading to isospin multiplets. 
In the restoration limit \eqref{eq:restore}, which leads to the  enhanced flavour symmetry $\SU(N_f)_V \to \SU(N_f)_V \times \SU(N_f)_A  \times \U(1)_A$,  the same types of degeneracy can be expected.
There is, however, another important point that accompanies this 
effect: namely, that the additional global 
symmetry gives rise to new quantum numbers. 

\begin{figure}[t]
 \includegraphics[scale=0.58,trim={0.5cm 3.5cm  0 0},clip]{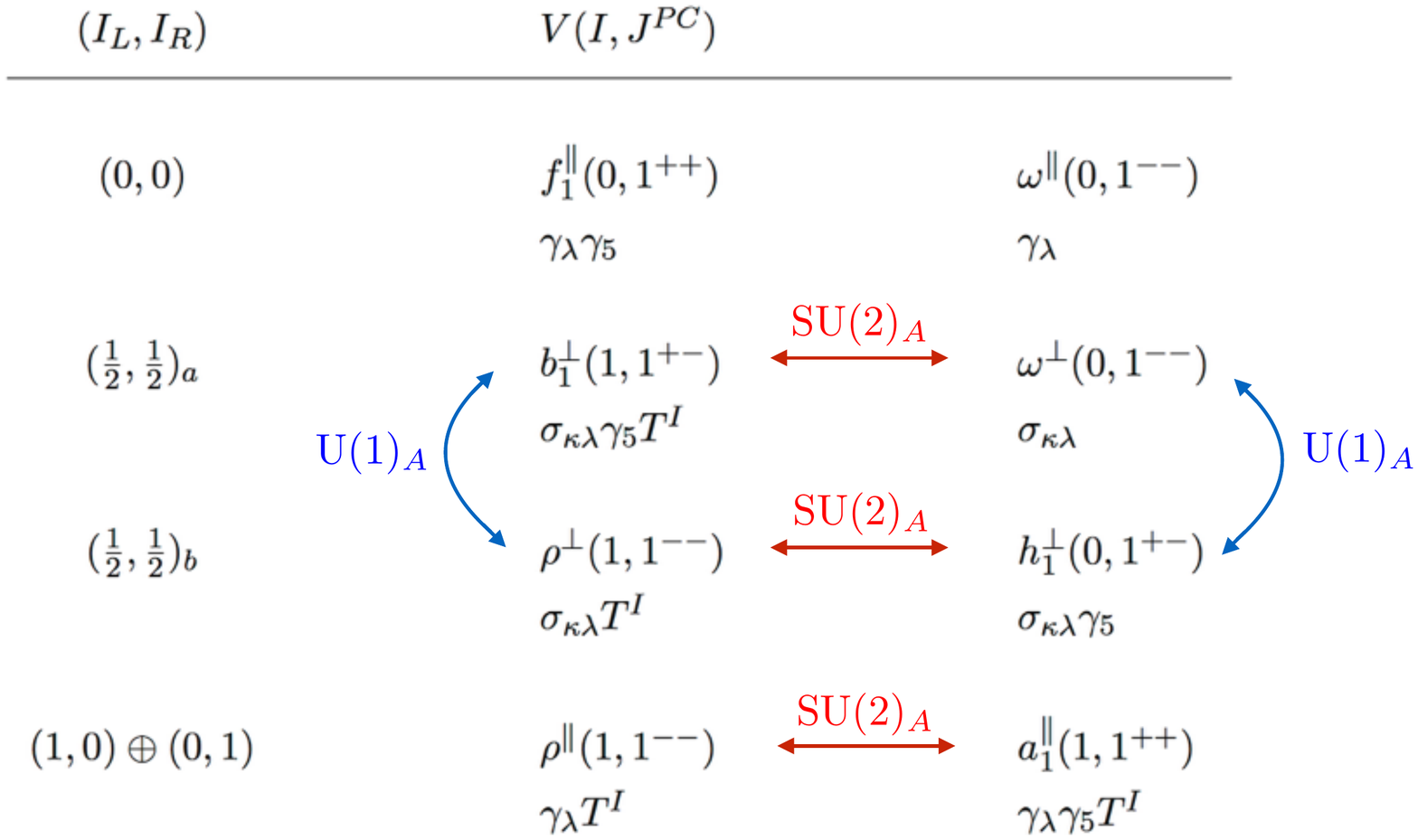}
\caption{\small Lowest-lying particles coupling to vector and tensor currents
of isosinglet- and isotriplet-type, making a total of $16$ vector mesons. 
For simple comparison, we use the same graphic presentation with similar notation as used in \cite{DGL14}.
The interpolating operators are defined in \eqref{eq:inter2} where the coupling to the states is specified 
further below.
In the restoration limit \eqref{eq:restore}, 
 one would expect a $4$-, $4$- and $6$-plet degeneracy  by the restoration of the $\SU(2)_A$ (red arrows). 
In the case where  $\U(1)_A$ (blue arrows) symmetry is also restored, this leads to a $8$- and $6$-plet degeneracy. The actual degeneracy is found to be larger \cite{DGL14}, compatible 
 with an emergent $\SU(4)$ \cite{SU4a} (or even $\SU(4)\times \SU(4)$ \cite{SU4SU4}) symmetry.}
\label{fig:degen}
\end{figure}

For the sake of concreteness, let us choose $N_f = 2$ below.
In the case at hand, $\SU(2)_V \times \SU(2)_A  \simeq \SU(2)_L \times \SU(2)_R$, this leads to both a left- and right-handed 
isospin quantum number 
$(I_L,I_R)$ instead of just the isospin $I_V$ itself.  
The classification of the representations is discussed in \cite{Cohen:1996sb}.
More precisely, the particles are classified according to the parity-chiral group 
$\SU(2) \times \SU(2) \times C_i$, where $C_i$ is the space reflection. 
The lowest  irreducible representations $(I_L,I_R)$, of dimension  $(2 I_L +1)(2 I_R +1)$, are
listed in figure \ref{fig:degen}. The splitting from left to right can be understood as the branching rule
of $(I_L,I_R)|_{\SU_V(2)}$; e.g. 
$(1/2,1/2)_a|_{\SU_V(2)} \to  3_{b_1^\perp} + 1_{\omega^\perp}$.  The two interpolating currents discussed as templates in the text \eqref{eq:inter2} 
correspond to the $(1,0) \oplus (0,1)$ multiplet. The two $(1/2,1/2)$ representations denoted by superscripts $a$ and $b$ are distinct by the parity operation.
The use of $\parallel$ and $\perp$ as superscripts is 
non-standard, and inspired by the notation of the corresponding decay constants. 
As a last generic remark, let us add that, in the real world, the $\rho$ and the $\rho' = \rho(\text{\small{1450}})$ are
admixtures of the $\rho^\parallel$ and $\rho^\perp$-states.
In the following subsection, we give an example where one can explicitly see how the
$\U(1)_A \times \SU(2)_A$-violating  condensates control differences in the hadronic 
data between the $\rho$ and $a_1$. 

\subsection{The Weinberg sum rules as an example}

From group-theoretic considerations, one can argue that the first correction to the 
chirality correlation function \eqref{eq:start} is given by
$(\Pi^{a,b}_{LR})_{\mu \nu}(q^2)   \sim  \vev{O_{\chi SB} } Q^{-6} + O(Q^{-8})$, 
where $q^2 \equiv - Q^2$ (e.g. \cite{Kapusta:1993hq}),  
$4 O_{\chi SB}  = \bar q T^3 \la^i  q_L \bar q T^3 \la^i  q_R $, and  $\la^i$  are the $\SU(3)$ colour matrices.   
One can then power-expand the denominator and arrive at 
the first two Weinberg sum rules 
\begin{equation}
\int_0^\infty s^n ( \rho_V(s) -  \rho_A(s)) = 0 \;, \quad n = 0 ,1 \;.
\end{equation}
The leading correction, or the third sum rule, is given by
\begin{equation}
\int_0^\infty s^2 ( \rho_V(s) -  \rho_A(s)) =  2 \pi \al_s \vev{O_{\chi SB} }   \;.
\end{equation}
So far, everything is exact.  Since the correlation functions are well-described at high $q^2$ by perturbation 
theory, which is equal for the $V$- and the $A$-channel, $\rho_V(s) \simeq \rho_A(s)$ will hold 
for some $s > s_0$. Weinberg \cite{WSR} assumed that $s_0$ is just above the $a_1$ resonance, and restricted 
himself to the parametrisations found in \eqref{eq:rho_app}, from which he deduced 
\begin{alignat}{3}
&  F_\rho^2 - F_\pi^2 - F_{a_1}^2 &\;=\;& 0  \;, \nonumber \\[0.1cm]  
& m_\rho^2 F_\rho^2-  m_{a_1}^2 F_{a_1}^2  &\;=\;& 0 \;, \nonumber \\[0.1cm]  
& m_\rho^4 F_\rho^2 - m_{a_1}^4 F_{a_1}^2   &\;=\;&   2 \pi \al_s \vev{O_{\chi SB} } \;.
\end{alignat}
These sum rules are  rather well-satisfied at the empirical level. The last equation nicely illustrates, in  a concrete setting, how the 
restoration limit is controlled by the $ \SU(2)_A  \times \U(1)_A $-violating condensate $\vev{O_{\chi SB} }$.
We note that,
in the vacuum factorisation approximation,
 $\vev{O_{\chi SB} } = - \frac{N_c^2-1}{ N_c^2}  \vev{\bar qq}^2 $ \cite{PT84}.

\section{Definition of effective Hamiltonian}
\label{app:effectiveH}

Here we describe in more detail the effective Hamiltonian for $b \to (d,s) \ga$ and $b \to (d,s) \ell \bar{\ell}$ decays, clarifying the notation in \eqref{eq:Heff}. In the basis of \cite{Buchalla:1995vs}, the operators contributing to \eqref{eq:Heff} are: the four-quark tree-level operators 
\begin{align}O^\UU_{1} &=  {\bar \DD }_{L,i}  \gamma_\mu \UU_j \bar \UU_{L,j}  \gamma^\mu  b_i \, , \nonumber \\  
 O^\UU_{2} &=  {\bar \DD }_L  \gamma_\mu  \UU \bar \UU_L  \gamma^\mu  b \, ;
 \end{align}
 the loop-induced four-quark operators, $O_{3,...,6}$:
 \begin{alignat}{2}
 O_3= \left(\bar{D}_{L}  \ga_\mu b \right) \sum_{q}\left( \bar{q}_{L} \ga^\mu q \right) \, ,  & \qquad \qquad
O_4= \left(\bar{D}_{L,i} \ga_\mu b_{j} \right) \sum_{q}\left( \bar{q}_{L,j} \ga^\mu q_{i} \right) \, , \nonumber \\
O_5= \left(\bar{D}_{L} \ga_\mu b \right) \sum_{q}\left( \bar{q}_{R}\ga^\mu q \right)\, ,  & \qquad \qquad
O_6= \left(\bar{D}_{L,i} \ga_\mu b_{j} \right) \sum_{q}\left( \bar{q}_{R,j}\ga^\mu q_{i} \right) \, ; 
 \end{alignat}
 and the electromagnetic and QCD dipoles (with $D_\mu = \partial_\mu - i g_s G_\mu- i e A_\mu $ convention)
 \begin{equation}
 O_{7(8)} =  -\frac{e(g)}{16 \pi^2}m_b  {\bar s}_L  \sigma \cdot F(G) b \, .
 \end{equation}
 The $C_i$ are scale-dependent Wilson coefficients, which can be calculated perturbatively using 
renormalisation group methods (e.g. \cite{Buchalla:1995vs,CMM1996}).
The most relevant $C_i$ for the discussion in the paper are 
$(C_2 ,C_1,C_7)(m_b) \simeq (1,-0.13,-0.37)$. 

For $b \to (d,s) \ell \bar{\ell}$ decays, one also needs the operators
\begin{equation}
O_{9,(10)} = \frac{\al_{\textrm{EM}}}{4\pi} {\bar \DD }_L  \gamma_\mu b \bar{\ell} \ga^\mu \left( \ga_5 \right) \ell \, .
\end{equation}
Parity-flipped versions of the operators above can be obtained by the replacement $O' = \left.O\right|_{{\bar \DD }_L \to {\bar \DD }_R}$, as well as (in the \ac{SM}) $m_b \to m_D$ in $O_{7,8}$.  
The relative importance of the $O_{1,2}^{U=u,c}$ operators is dependent on the CKM hierarchy 
$\lambda_{\UU }^{(\DD)} = V_{ \UU b}^{\vphantom{*}}V_{\UU \DD}^*$  for $b \to d$ and $b \to s$
 \begin{equation}
 \label{eq:Wolf}
 \la^{(s)}_u : \la^{(s)}_c : \la^{(s)}_t = \lambda^4:\lambda^2:\lambda^2  \;, \qquad
 \la^{(d)}_u : \la^{(d)}_c : \la^{(d)}_t = \lambda^3:\lambda^3:\lambda^3 \;,
\end{equation}
respectively. Above,  $\la$ is the Wolfenstein parameter, with the approximate value 
$ \la \simeq 0.225$ \cite{PDG2016}.

\bibliographystyle{utphys}
\bibliography{References_All}

\end{document}